\documentclass[prb,twocolumn,superscriptaddress,showpacs,a4paper]{revtex4}
\usepackage{times} \normalfont \usepackage[T1]{fontenc}
\usepackage{graphicx} % \usepackage{color}

\def\ealla#1{{\rm e}^{#1}}
\def\pH2{$p$H$_2$}
  
\begin{document}

\title{
  Computer simulation of quantum melting in hydrogen clusters
}

\author{Stefano Baroni}
\affiliation{
  SISSA -- Scuola Internazionale Superiore di Studi Avanzati \\
  and INFM {\sl DEMOCRITOS} National  Simulation Center \\ Via
  Beirut 2-4, I-34014 Trieste, Italy
}
\author{Saverio Moroni}
\affiliation{
  INFM {\sl SMC} National Research Center \\
  and Dipartimento di Fisica, Universit\`a di Roma
  {\sl La Sapienza} \\ 
  Piazzale Aldo Moro 2, I-00185 Rome, Italy
}

\begin{abstract}
  We introduce a new criterion---based on multipole dynamical
  correlations calculated within {\em Reptation Quantum Monte
    Carlo}---to discriminate between a melting {\em vs.} freezing
  behavior in quantum clusters. This criterion is applied to small
  clusters of para-hydrogen molecules (both pristine and doped with a
  CO cromophore), for cluster sizes around 12 molecules. This is a
  {\em magic} size at which para-hydrogen clusters display an
  icosahedral structure and a large stability. In spite of the similar
  geometric structure of CO@(\pH2)$_{12}$ and (\pH2)$_{13}$, the first
  system has a rigid, {\em crystalline}, behavior, while the second
  behaves more like a {\em superfluid} (or, possibly, a {\em
    supersolid}).
\end{abstract}


\maketitle

%36.40.-c Atomic and molecular clusters
%34.30.+h Intramolecular energy transfer; intramolecular dynamics; 
%         dynamics of van der Waals molecules
%67.40.Yv Impurities and other defects
%36.40.Mr Spectroscopy and geometrical structure of clusters
%02.70.Ss Quantum Monte Carlo methods
%61.46.+w Nanoscale materials: clusters, nanoparticles, nanotubes, 
%         and nanocrystal 
%33.20.Sn Rotational analysis
 
Understanding the dynamics of quantum many-body systems is one
of the major challenges presently set to theoretical and computational
condensed-matter physicists. The combination of density-functional
theory with molecular dynamics realized 20 years ago by Car and
Parrinello \cite{CarParr} opened the way to the study of the dynamics of
quantum driven classical systems ({\em i.e.} of atomic systems whose
dynamics is essentially classical, but driven by quantum-mechanical
forces). The development of density-functional perturbation theory has
allowed for a systematic calculation of the low-lying quantum excited
states of these same systems in the harmonic approximation \cite{BGT}.
Quantum Monte Carlo methods, on the other hand, have been very
successful in describing the ground-state and finite-temperature
properties of interacting bosons \cite{RMP-Ceperley} and of lattice
models of strongly interacting fermions \cite{fermion-lattice}, and
they promise a similar success in the study of chemical systems in the
near future \cite{QMC-RMP}. In spite of all these progresses, the
ability to calculate in a reliable way the properties of the excited
states of interacting quantum systems remains a largely unachieved
goal. Recent advances in the quantum Monte Carlo methodology have
partially changed this scenario, at least in what concerns systems
whose ground-state wave-function is positive ({\em i.e.} bosons) and
whose low-lying energy spectrum is dominated by few excited states,
which is the typical situation for superfluids \cite{noantri,genova}.
Thanks to these advances, it is now possible to calculate the
low-lying excitation spectrum of clusters of up to a few tens $^4$He
atoms, possibly doped with some cromophore molecules which are
experimentally used as spectroscopic probes of the dynamical and
superfluid properties of the droplet.

Molecular hydrogen, in its nuclear-spin realization called {\em
  para-hydrogen} (\pH2, $I=J=0$), is the only substance occurring
in nature, other than $^4$He, which can possibly exhibit the
phenomenon of superfluidity at (not too) low temperature \cite{ginzburg}. In
spite of the lighter mass with respect to $^4$He, the intermolecular
potential is so much stronger that the estimated value of the
$\lambda$-transition temperature ($\approx 2\rm ^\circ K$) is much
lower than the observed triple-point temperature ($13.96\rm ^\circ
K$). For this reason, much attention has been and is being paid to
those confined geometries (such as clusters
\cite{ceperley_h2,toennies-science,levi,whaley-h2,toennies-PRL} or
films \cite{gordillo-ceperley,boninsegni}) which may hinder
crystallization and make superfluidity (or what remains of it in these
geometries) observable in \pH2. Discriminating between a melting
{\em vs.} freezing behavior in a finite system is a subtle issue
which requires a careful consideration of the dynamical evolution of
the system.  In this paper we propose a strategy to cope with this
problem, based on {\em Reptation Quantum Monte Carlo} (RQMC), a method
that we developed a while ago to deal with dynamical correlations in
systems of interacting bosons \cite{noantri}. As a (still very
preliminary) application, we show how using suitable
(imaginary-) time correlation functions it is easy (and fun!) to
discriminate between the melting {\em vs.} freezing behavior of
small \pH2 clusters---with or without a CO molecule solvated in
them---for sizes close to $n=13$, a magic number which favors
crystallization.

\section*{Reptation Quantum Monte Carlo}

The dynamics of classical stochastic processes bears close
similarities with the time evolution of quantum systems in imaginary
time. These similarities---together with the fact that every system
tends towards the ground state at large imaginary times---lay at the
basis of many quantum Monte Carlo methods which sample ground state
properties from the equilibrium properties of a suitably defined
classical random walk. Comparatively minor attention has been paid to
the dynamical properties of the random walks used in quantum
simulations. In this section we show how a careful
exploitation of these similarities can be used to estimate the
dynamical properties of strongly interacting boson fluids.

Most ground-state Monte Carlo techniques are based on the property
that the imaginary-time evolution of (almost) any trial wave-function,
$\Phi_0$, in the infinite-time limit tends to the ground state,
$\Psi_0$. As a consequence, the ground-state energy can
be expressed as:
\begin{eqnarray} 
  E_0 &=& \lim_{\tau\to\infty} {\langle \Phi_0 | H {\rm e}^{-H\tau} 
    | \Phi_0 \rangle \over \langle \Phi_0 | {\rm e}^{-H\tau}
    | \Phi_0 \rangle} 
  \label{eq:ground-state-1} \\
  &=& -\lim_{\tau\to\infty} {d\over d\tau}\log{\cal Z}_0 
  \label{eq:ground-state-F} \\
  &=& -\lim_{\tau\to\infty} {1\over \tau}\log{\cal Z}_0, 
  \label{eq:ground-state-E} 
\end{eqnarray}
where $H = -{\partial^2 \over \partial x^2} + V(x)$ is the Hamiltonian
of the system and ${\cal Z}_0 = \langle \Phi_0 | {\rm e}^{-H\tau} |
\Phi_0 \rangle$. If ${\cal Z}_0$ is thought as the partition function
of an auxiliary, fictitious, system, then Eqs.
(\ref{eq:ground-state-F}) and (\ref{eq:ground-state-E}) express the
free and internal energies of this system in the zero-temperature
limit. By using Eq.  (\ref{eq:ground-state-E}), ground-state
expectation values of static operators and generalized
susceptibilities can be expressed as first and second logarithmic
derivatives of ${\cal Z}_0$ with respect to the strength of a suitably
chosen perturbation to the Hamiltonian \cite{noantri}.

A link with the theory of stochastic processes can be established by
splitting the Hamiltonian as:
\begin{equation}
  H = {\cal H} + {\cal E}(x),
\end{equation}
where ${\cal H} = -{\partial^2 \over \partial x^2} + {1\over
  \Phi_0(x)} \Phi''_0(x)$, ${\cal E}(x) = -{1\over \Phi_0(x)}
\Phi''_0(x) + V(x)$, and the double prime indicates the Laplacian.
Note that by construction $\Phi_0$ is an eigenstate of $\cal H$ with
zero eigenvaue and that, if it is chosen to be node-less, it also has
to be its non-degenerate ground state, so that the excited-state
spectrum of $\cal H$ is strictly positive. ${\cal E}(x)$ is easily
recognized as the definition of the {\em local energy}, well known to
the variational Monte Carlo (VMC) and diffusion Monte Carlo (DMC)
practitioners. The pseudo-partition function, ${\cal Z}_0$, can be
given a path-integral representation by Trotter-splicing the
propagator appearing therein:
\refstepcounter{equation}\label{eq:Z0}
$$\displaylines{
  \quad
  {\cal Z}_0 = \int \Phi_0(x_N) 
  {\cal G}_\epsilon(x_N,x_{N-1})
  {\cal G}_\epsilon(x_{N-1},x_{N-2})
  \cdots \hfill \cr \hfill
  {\cal G}_\epsilon(x_1,x_0) \Phi_0(x_0) 
  {\rm e}^{-\epsilon \sum_n {\cal E}(x_n)}
  {\cal D}[X]. 
\quad (\theequation)
}$$
The short-time propagator, ${\cal G}_\epsilon$, can be expressed in
terms of the transition matrix of a suitably defined Markovian random
walk:
\begin{equation}
  {\cal G}_\epsilon(y,x) \propto {1\over \Phi_0(x)} {\cal
    W}_\epsilon(y,x) \Phi_0(y),
  \label{eq:GvsW}
\end{equation}
where $ {\cal W}_\epsilon(y,x) \propto {\rm
  e}^{-{(y-x-\epsilon f(x))^2 \over 4\epsilon}} $ is the transition
probability associated with the Langevin equation:
\begin{equation} 
  dx = f(x) \epsilon + d\xi, \label{eq:langevin} 
\end{equation} 
$f(x) = -{1\over \Phi_0(x)}{\partial \Phi_0(x) \over \partial x}$, and
$\xi(\tau)$ is a Wiener process: $\langle d\xi\rangle = 0; \quad
\langle (d\xi)^2 \rangle=2\epsilon$. $P_0(x) = \Phi_0(x)^2$ is the
(unique) equilibrium distribution of Eq. (\ref{eq:langevin}). This
fact, together with Eq. (\ref{eq:GvsW}), allows to cast
Eq. (\ref{eq:Z0}) into the form:
\begin{equation}
  {\cal Z}_0 = 
  \left\langle
    {\rm e}^{-{\cal S}[X]}
  \right\rangle_{RW},
\end{equation}
where ${\cal S}[X]=\epsilon\sum_n{\cal E}(x_n)$ is the {\em reduced
  action} of the system, and the brackets, $\langle\cdot\rangle_{RW}$,
indicate a statistical average over an equilibrium distribution of
quantum paths (or random walks), $\{X\}$.

Suppose now that the system is coupled to a (imaginary-) time
dependent perturbation: $H_\lambda = H + \lambda(\tau) A$. The second
derivatives of the pseudo-partition function, ${\cal Z}_0$, with
respect to $\lambda$ calculated at $\lambda=0$ give the ground-state
imaginary-time correlations of the $A$ operator:
\stepcounter{equation}
$$\displaylines{ \quad \langle A(t+\tau) A(t) \rangle \hfill \cr
  \quad\quad\quad \quad\quad = { \langle \Phi_0 | \ealla{-(T-t-\tau)H}
    A \ealla{-\tau H} A \ealla{-tH} | \Phi_0 \rangle \over \langle
    \Phi_0 | \ealla{-TH}| \Phi_0 \rangle } \hfill\cr \quad\quad\quad
  \quad\quad = { \left\langle A(t+\tau) A(t) {\rm e}^{-{\cal S}[X]}
    \right\rangle_{RW} \over \left\langle \ealla{-{\cal S}[X]}
    \right\rangle_{RW} }. \hfill (\theequation) }$$
According to the
above equations, a practical algorithm to calculate ground-state time
correlations would be to generate segments of a random walk according
to the Langevin equation, Eq.  (\ref{eq:langevin}); each segment is
then accepted or rejected according to a Metropolis test performed on the
reduced action, ${\cal S}[X]$; the statistical average of time
correlations calculated along the paths thus generated will provide
the desired quantum time correlations. This is the heart of the {\em
  Reptation Quantum Monte Carlo} method \cite{noantri}. This algorithm
can be generalized by generating trial moves for the paths according
to any a-priori probability distribution, and then accepting or
rejecting them according to a Metropolis test done one the actual
probability distribution for the paths, Eq. (\ref{eq:Z0}),
\cite{PIGS}.

The time correlation function of various observables which couple to
external fields (such as, {\em e.g.}, the electric dipole) are the
Laplace transforms of spectral weight functions which, in some
favorable cases, can be directly compared with experiments
\cite{genova}. In the next section we show how suitably defined
time-correlation functions can be used to discriminate the quantum
melting {\em vs.} freezing behavior of small \pH2 clusters. 

\section*{Magic \pH2 clusters}

Infrared and microwave spectroscopies of solvated cromophores are
being widely used to probe the microscopic structure of $^4$He
nano-droplets, as well as their propensity to display a superfluid
behavior \cite{he-spectroscopy}. It is hoped that the use of these
techniques for \pH2 clusters of different size will help identify
fingerprints of superfluidity, and to clarify the interplay between
this phenomenon and the competing tendency to crystallization in these
systems. Infrared spectra spectra of carbon mono-oxide (CO) molecules
solvated in small (\pH2)$_n$ clusters ($n\le 14$) have been
recently measured and RQMC simulation have helped unravel them and
assign the identity of the individual lines \cite{botti}. 

\begin{figure}% [float]
  \begin{center}
    \includegraphics[width=9cm]{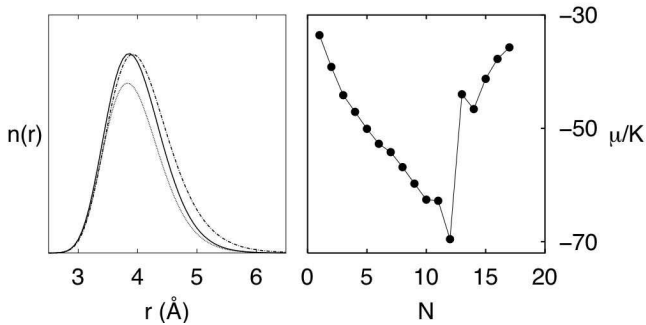} 
    \vspace{-10mm}
  \end{center}
  \caption{Left panel: radial density distribution function of \pH2
    molecules in CO@(\pH2)$_n$ (dotted line: $n=11$, continuous line:
    $n=12$; dash-dotted line: $n=13$). Right panel: \pH2 chemical
    potential ($\mu_N=E_n-E_{n-1}$) as a function of the cluster size
    in CO@(\pH2)$_n$. 
  }
\end{figure}

In Fig. 1 we display the radial distribution functions of \pH2 around
a CO molecule, for cluster sizes around the {\em magic} number $n=12$
\cite{botti}. For $n\le 12$ the magnitude of the maximum of these
functions increases with the cluster size. For $n\ge 12$ this value
stays nearly constant and the increase of the cluster size shows in
the tail of the distribution, rather than in the height of the peak,
indicating that the first solvation shell is completed at this magic
number. For $n=12$ the absolute value of chemical potential also
displays a maximum, 
suggesting the greatest stability of the clusters at this size. At the
same time, the rotational distortion constant displays a minimum,
indicating the largest rigidity of this clusters.  All these findings
suggest that the propensity of the \pH2 droplet to crystallize is
maximum ad this cluster size. This fact has to be compared with the
findings of Ref.  \cite{ceperley_h2} which show clear signs of
superfluidity in a cluster of 13 \pH2 molecules, whose structure is
similar to that of CO@(\pH2)$_n$ with the solvated CO molecule
substituted with a \pH2 molecule. In order to assess the melting {\em
  vs.} freezing behavior of finite systems, we introduce a
dynamical criterion based on the persistence of geometrical signatures
of the {\em crystal structure}.

The shape of a rigid body is well described by the multipole moments
of, say, its mass density distribution around the center of mass,
$Q^l_m$. In order to characterize the degree of rigidity of a cluster
which, in addition to shape fluctuations, also experiences rotational
diffusion, the shape must be described in a way which is independent
of the orientation. The magnitude of the multipole, defined as $q^l =
\sum_m {Q^l_m}^* Q^l_m$, provides such an invariant characterization
of a rigid body. This quantity, however, is not suitable to
discriminate the cases where a given multipole vanishes on the
average, and $\langle q^l\rangle \ne 0$ because of the fluctuations,
from those where a non vanishing value of $\langle q^l\rangle$ is due
to the average shape of the cluster. This discrimination is best
achieved in terms of a suitably defined rotationally-invariant time
correlation functions of the various multipole moments in the
{\em rotating-axes} frame. In order to proceed, let us define an
effective angular 
velocity, $\mathbf{\Omega}(\tau)$, as the average of the forward
angular velocities of the individual molecules. The operator
corresponding to the global rotation accomplished in a time step
$\epsilon$ reads:
\begin{equation}
\Delta R(\tau) = \ealla{-i \epsilon {\bf L} \cdot \mathbf{\Omega}(\tau) },
\end{equation}
where $\bf L$ is the generator of the rotation group (angular
momentum). The total rotation accomplished  during a time $\tau$ is:
 \begin{eqnarray}
  R(\tau) &=& {\cal T}\ealla{-i\int_0^\tau  {\bf L} \cdot
  \mathbf{\Omega}(\tau') d\tau'} \\ 
  &\equiv& \prod_{n=1}^{T/\epsilon} \Delta R(n\epsilon),
  \label{eq:Rtot}
\end{eqnarray} 
where ${\cal T}\ealla{(\cdot)}$ indicates the time-ordered
exponential. With the aid $R(\tau)$ we define a {\em
  rotating-axes} multipole time-correlation function, $c_l(\tau)$, as the
correlation between the multipole moment at (imaginary) time $\tau'$
and the multipole at time $\tau'+\tau$, {\em rotated by $
  R(\tau')^{-1}$} (time-translation invariance implies that these
correlations are independent of $\tau'$):
\begin{equation}
  c_l(T) = {\sum_{m} \left\langle Q^{l*}_m(\tau) \bar{Q}^l_m(\tau+T) 
    \right \rangle \over \sum_{m} \left\langle Q^{l*}_m(\tau)
      Q^l_m(\tau) \right \rangle},
  \label{eq:che_le_palle_ancor_gli_girano} 
\end{equation}
where $Q^l_m(\tau)$ is the multipole moment of the system at
(imaginary) time $\tau$, the brackets, $\langle\cdot\rangle$, indicate
an average over the random walk, and $\bar Q$ indicates the multipole
{\em rotated by $R^{-1}(\tau)$}. If a system is rigid, $c_l(\tau)$ is
independent of time, whereas it tends to zero at large values of
$\tau$ whenever the shape of the system undergoes random fluctuations.
A large value of $c_l(\tau)$ at large times is thus an indicator of the
persistence of the shape of the system \cite{paolo}.

In the case of CO@(\pH2)$_{12}$ our simulations indicate a very
rapid decay of the correlations for all the multipoles up to and
including $l=5$. For $l=6$, instead, after a short transient in which
a very steep drop occurs, the correlations decay very slowly,
indicating a long persistence of a shape compatible with a
nonvanishing $l=6$ multipole and vanishing multipoles with smaller
values of $l$. In Fig. 2 we report the $l=6$ {\em rotating-axes}
multipole time correlation function calculated for CO@(\pH2)$_{12}$.

\begin{figure}[h]
  \begin{center}
    \vspace{-5mm}
    \includegraphics[width=9cm]{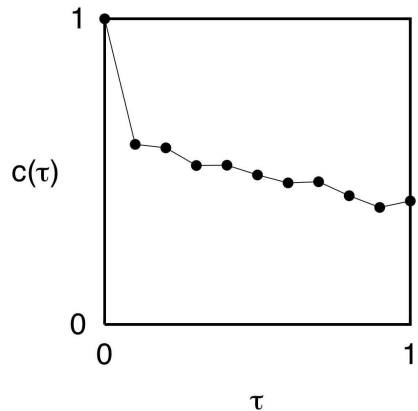} 
    \vspace{-15mm}
  \end{center}
  \caption{
   $l=6$ {\em rotating-axes} multipole time correlation function
   (Eq. \ref{eq:che_le_palle_ancor_gli_girano}) for CO@(\pH2)$_{12}$. 
  }
\end{figure}

This behavior is indeed compatible with a regular icosahedral shape,
whose lowest non-vanishing multipoles correspond to $l=6$ and $l=10$.
In order to visualize the intrinsic shape of the cluster, we calculate
the {\em rotating-axes} number density distribution of the hydrogen
molecule, $\bar n_{\rm H_2}({\bf r})$, by rotating each configuration
of a quantum path at imaginary time $\tau$ with respect to the first configuration
by $R^{-1}(\tau)$, as
defined in Eqs. (\ref{eq:Rtot}). The contribution to the density from
each quantum path sampled by our RQMC procedure is then further
rotated, so as to minimize the mutual distance of the particle
centroids resulting from different paths. In Fig. 3 we display the
{\em rotating-axes} number densiy distribution resulting from our
simulations for CO@(\pH2)$_{12}$.

\begin{figure}% [float]
  \begin{center}
    \includegraphics[width=7cm]{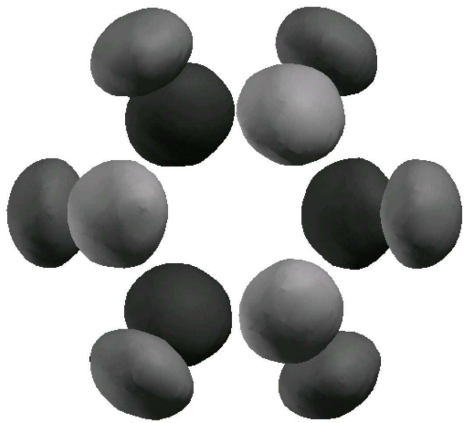} 
  \end{center}
  \caption{
    \pH2 {\em rotating-axes} number-density distribution for
    CO@(\pH2)$_{12}$ 
    (see text). The displayed iso-surface corresponds to a value of
    the density including 60\% of the total number of H$_2$
    molecules. The solvated CO molecule rotates nearly freely
    with respect to the {\em rotating-axes} frame, and it is not
    displayed in the figure.
  }
\end{figure}

The results displayed in Figs. 2 and 3 seem to indicate that
CO@(\pH2)$_{12}$ clusters behave much as rigid bodies, the closest
realization of a {\em crystal} in a finite system. This behavior is
different from that found in Ref. \cite{ceperley_h2} for undoped
(\pH2)$_{13}$ clusters whose structure is expected to be very similar
to that of CO@(\pH2)$_{12}$, the main difference being the
substitution of the central CO molecule with another \pH2 molecule. In
that paper the behavior of (\pH2)$_{13}$ was described as that of a
{\em structured superfluid}, possibly resembling a {\em supersolid}
\cite{ceperley_h2}. The prediction of a liquid-like behavior
  for (\pH2)$_{13}$---as well as its enhanced stability with respect
  to small clusters of different size---is supported by recent Raman
  spectroscopy measurements of cryogenic (\pH2)$_n$ free jets
  \cite{toennies-PRL}. Diffusion quantum Monte Carlo simulations
  presented in that paper were used to assign individual vibrational
  lines to clusters of different size, and they support the
  delocalized character of these systems, as well as the greatest
  stability of (\pH2)$_{13}$. 

\begin{figure}[h]
  \begin{center}
    \includegraphics[width=9cm]{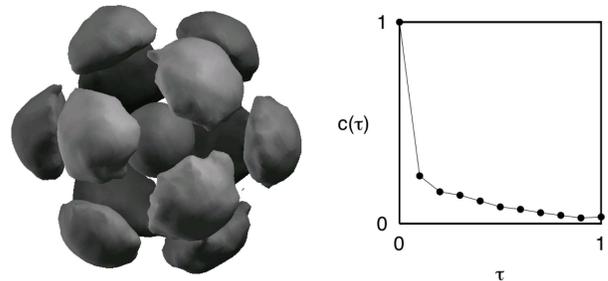} 
  \end{center}
  \caption{
    Left panel: \pH2 {\em rotating-axes number density distribution} for
    (\pH2)$_{13}$ (see text). The displayed iso-surface corresponds
    to a value of the density including 60\% of the total number of
    \pH2 molecules. Right panel: $l=6$ {\em rotating-axes} multipole time
    correlation function (Eq. \ref{eq:che_le_palle_ancor_gli_girano})
    for the same system.
  }
\end{figure}

In Fig. 4 we report the {\em rotating-axes} number-density
distribution and the $l=6$ {\em rotating-axes} multipole time
correlation function, resulting from RQMC simulations for
(\pH2)$_{13}$. Although the number density distribution displays a
nicely icosahedral shape---as it is the case for
CO@(\pH2)$_{12}$---the $l=6$ multipole correlation function displays a
behavior which---as it will be shown below---is typical of a molten
cluster. We interpret this seemingly contradictory behavior as a
consequence of higher frequency of quantum inter-molecular exchanges
in (\pH2)$_{13}$ than in CO@(\pH2)$_{12}$.  This is due to both the
lack of exchanges between molecules in the outer shell and the central
molecule in CO@(\pH2)$_{12}$, and also to a larger binding of the
molecules in the outer shell to the central one, which acts as to
hinder intra-shell exchanges. During an exchange the shape of the
cluster strongly departs from icosahedral. We expect therefore
that---when the ratio between the typical duration of an exchange
process and the waiting time between two exchanges increases---the
$l=6$ multipole autocorrelations would decrease rapidly. It is
interesting to notice that---contrary to CO@(\pH2)$_{12}$, where the
decay of the $l=6$ autocorrelations is much slower than for smaller
angular momenta---in the case of (\pH2)$_{13}$ autocorrelations of
different multipoles decay in a qualitatively similar way.

\begin{figure}[h]
  \begin{center}
    \includegraphics[width=9cm]{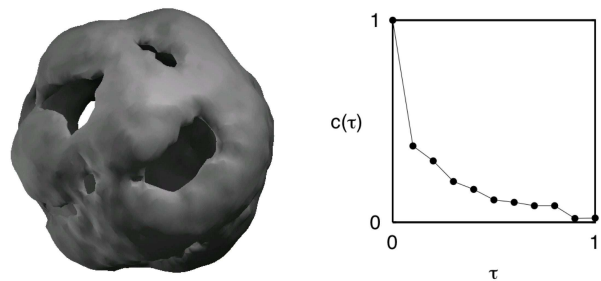} 
  \end{center}
  \caption{
    Left panel: \pH2 {\em rotating-axes} number density distribution for
    CO@(\pH2)$_{13}$ (see text). The displayed iso-surface corresponds
    to a value of the density including 60\% of the total number of
    \pH2 molecules. Right panel: $l=6$ {\em rotating-axes} multipole time
    correlation function (Eq. \ref{eq:che_le_palle_ancor_gli_girano})
    for the same system.
  }
\end{figure}

In Fig. 5 we report the {\em rotating-axes} number density
distribution and the $l=6$ {\em rotating-axes} multipole time
correlation function, resulting from RQMC simulations for
CO@(H$_2$)$_{13}$. The shape of the number-density iso-surface is
strongly irregular, as a consequence of the quantum exchanges between
one of the \pH2 molecules in the first solvation shell and the
additional (thirteenth) molecule at the center of the cluster. The
$l=6$ multipole time 
correlations correspondigly display a rather fast decay. The decay of
the these correlations is in fact slower than for (H$_2$)$_{13}$ for
which, however, the shape of the number-density isosurface is much
more regular (only somewhat broadened with respect to
CO@(H$_2$)$_{12}$ which we would qualify as a {\it solid cluster}).

\section*{Conclusions}

We understand that our results---while confirming much of what is
already known about the quantum behavior of \pH2 clusters
\cite{ceperley_h2,toennies-PRL,botti} and their greater
propensity to crystallize when they are {\em seeded} by a foreign
molecule \cite{levi}---probably raise more questions than they answer.
A full understanding of the melting {\em vs.} freezing behavior in
these exotic, yet very interesting, quantum systems will require much
more work than it has been possible to report in this paper.

% \section*{Acknowledgements}

\begin{acknowledgements}
  We wish to thank Giacinto Scoles for sharing with us some of his
  knowledge in the field of quantum clusters and for his contagious
  enthusiasm for Science. We are also grateful to Stefano Fantoni for
  encouraging our interest in quantum fluids and for collaborating
  with us in this field. Finally, we thank our friend Paolo Giaquinta
  for informing us of his old work on the local structure of classical
  fluids \cite{paolo} and for pointing to us the analogies between his
  approach and ours.
\end{acknowledgements}


\begin{thebibliography}{99}
  
\bibitem{CarParr}
  R. Car and M. Parrinello, Phys. Rev. Lett. {\bf 55}, 2471 (1985).
  
\bibitem{BGT}
  S. Baroni, P. Giannozzi, and A. Testa,  Phys. Rev. Lett. {\bf 58},
  1861 (1987). 
  
\bibitem{RMP-Ceperley}
  D. M. Ceperley, Rev. Mod. Phys. 67, 279 355 (1995) and references
  quoted therein.
  
\bibitem{fermion-lattice}
 See, {\em e.g.}: {\em Monte Carlo method in the physical sciences},
 AIP Conference proceedings. Vol.690, edited by J.E. Gubernatis
 (Melville: American Institute of Physics).
  
\bibitem{QMC-RMP}
  W.M.C. Foulkes, L. Mitas, R.J. Needs, and G. Rajagopal,
  Rev. Mod. Phys. 73, 33 (2001).
  
\bibitem{noantri} 
  S. Baroni and S. Moroni, Phys. Rev. Lett. {\bf 82}, 4745 (1999), and
  in {\sl Quantum Monte Carlo Methods in Physics and Chemistry},
  edited by P. Nightingale and C.J. Umrigar.  NATO ASI Series, Series
  C, Mathematical and Physical Sciences, Vol.  525, (Kluwer Academic
  Publishers, Boston, 1999), p. 313 (see also: {\tt
    http://xxx.lanl.gov/abs/cond-mat/9808213}).

\bibitem{genova} 
  S. Moroni and S. Baroni, Proceedings of the {\em Conference on
  Computational Physics 2004}, edited by M. Ferrario, C. Pierleoni,
  and S. Melchionna, Comp. Phys. Comm. in press (2005).
  
\bibitem{ginzburg} 
  V.L. Ginzburg and A.A. Sobyanon, JETP Lett. {\bf 15}, 242 (1972). 
  
\bibitem{ceperley_h2}
  P. Sindzingre, D.M. Ceperley, and M.L. Klein, Phys. Rev. Lett.
  {\bf 67}, 1871 (1991).

\bibitem{toennies-science}
  S. Grebenev, B. Sartakov, J.P. Toennies, and A.F. Vilesov, Science 
  {\bf 289}, 1532 (2000).
  
\bibitem{levi}
  A.C. Levi and R. Mazzarello, J. Low Temp. Phys. {\bf 122}, 75
  (2001); R. Mazzarello and A.C. Levi, J. Low Temp. Phys. {\bf 127},
  259 (2002).

\bibitem{whaley-h2}
  Y. Kwon and K.B. Whaley, Phys. Rev. Lett. {\bf 89}, 273401 (2002).

\bibitem{toennies-PRL}
  G. Tejeda, J.M. Fern\'andez, S. Montero, D. Blume, and
  J.P. Toennies, Phys. Rev. Lett. {\bf 92}, 223401 (2004).

\bibitem{gordillo-ceperley}
  M.C. Gordillo and D.M. Ceperley, Phys. Rev. Lett. {\bf 79}, 3010
  (1997).

\bibitem{boninsegni}
  M. Boninsegni, Phys. Rev. B {\bf 70}, 125405 (2004).

\bibitem{PIGS}
  A. Sarsa, K.E. Schmidt and W.R. Magro, J. Chem. Phys. {\bf 113},
  1366 (2000).
  
\bibitem{he-spectroscopy}
  See the {\em special topic} issue on He nanodroplets, J. Chem. Phys. {\bf
  115}, 10065-10281 (2001).

\bibitem{botti}
  S. Moroni, M. Botti, S. de Palo, and A.R.W McKellar, J. Chem. Phys.
  in press (2005).
  
\bibitem{paolo}
  After this paper was completed and submitted, we learnt from Paolo
  Giaquinta that ideas similar to ours had been developed by him long
  ago to study the local structure of classical fluids (P. Giaquinta, CECAM
  internal report (Paris, 1978), unpublished).
\end{thebibliography}
\end{document}